\documentclass[]{revtex4}
\usepackage{graphicx}

\begin{document}

\title{High energy cosmic ray \& neutrino astronomy}
\author{E. Waxman}
\affiliation{Particle Phys. \& Astrophys. Dept., Weizmann Inst. of Science, Rehovot, Israel}
%\institute{E. Waxman \at Particle Phys. \& Astrophys. Dept., Weizmann Inst. of Science, Rehovot, Israel, \email{eli.waxman@weizmann.ac.il}}

%\tableofcontents
%\cleardoublepage
%\pagenumbering{arabic}

\begin{abstract}
Cosmic-rays with energies exceeding $10^{19}$~eV are referred to as Ultra High Energy Cosmic Rays (UHECRs). The sources of these particles and their acceleration mechanism are unknown, and for many years have been the issue of much debate. The first part of this review describes the main constraints, that are implied by UHECR observations on the properties of candidate UHECR sources, the candidate sources, and the related main open questions.\\
In order to address the challenges of identifying the UHECR sources and of probing the physical mechanisms driving them, a ``multi-messenger" approach will most likely be required, combining electromagnetic, cosmic-ray and neutrino observations. The second part of the review is devoted to a discussion of high energy neutrino astronomy. It is shown that detectors, which are currently under construction, are expected to reach the effective mass required for the detection of high energy extra-Galactic neutrino sources, and may therefore play a key role in the near future in resolving the main open questions. The detection of high energy neutrinos from extra-Galactic sources will not only provide constraints on the identity and underlying physics of UHECR sources, but may furthermore provide information on fundamental neutrino properties.
\end{abstract}
\maketitle

\section{Introduction}

Cosmic-rays (CRs) with energies exceeding $\sim10^{19}$~eV are referred to as Ultra High Energy Cosmic Rays (UHECRs). The sources of these particles, which are probably extra-Galactic, and their acceleration mechanism are unknown, and for many years have been the issue of much debate (e.g. \cite{Hillas,Bhattacharjee00,Nagano00,Berezinsky08} and references therein). The first part of this chapter, \S~\ref{sec:UHECR}, describes the main constraints that are implied by UHECR observations on the properties of candidate UHECR sources. The constraints derived under the assumption that UHECRs are protons, which is supported by most observations but questioned by some (see \S~\ref{ssec:primaries}--\S~\ref{ssec:aniso}, \S~\ref{sec:multi-messenger}), are summarized in \S~\ref{ssec:source_constraints}. In \S~\ref{ssec:predictions} it is shown that GRBs are the only known type of sources that satisfy these constraints. Testable predictions for the spectrum and arrival direction distribution of UHECRs, made by the GRB model of UHECR production, are also described.

The challenges of identifying the UHECR sources, and of probing the physical mechanisms driving them, may be met with the help of high energy neutrino detectors \cite{HalzenHooper02,Waxman05,Hoffman_nu_tel_09,Anchordoqui09}. This is discussed in \S~\ref{sec:nu}. In \S~\ref{ssec:bound} it is shown that detectors, which are currently under construction, are expected to reach the effective mass required for the detection of high energy extra-Galactic neutrino sources (see Figs.~\ref{fig:bound}, \ref{fig:GRB_nu}), and may therefore play a key role in the near future in resolving the main open questions. GZK and GRB neutrinos are discussed in \S~\ref{ssec:GZK} and \S~\ref{ssec:grb_nu} respectively. In \S~\ref{ssec:nu_phys} we point out that the detection of high energy neutrinos from extra-Galactic sources will not only provide constraints on the identity and underlying physics of UHECR sources, but may furthermore provide information on fundamental neutrino properties.

The main open questions associated with the production of UHECRs are summarized in \S~\ref{sec:multi-messenger}. It is argued that a ``multi-messenger" approach, combining electromagnetic, cosmic-ray and neutrino data, would be required in order to provide answers to these questions.

\section{What we (don't) know about the sources of UHECRs}
\label{sec:UHECR}

\begin{figure}
\includegraphics[width=12cm]{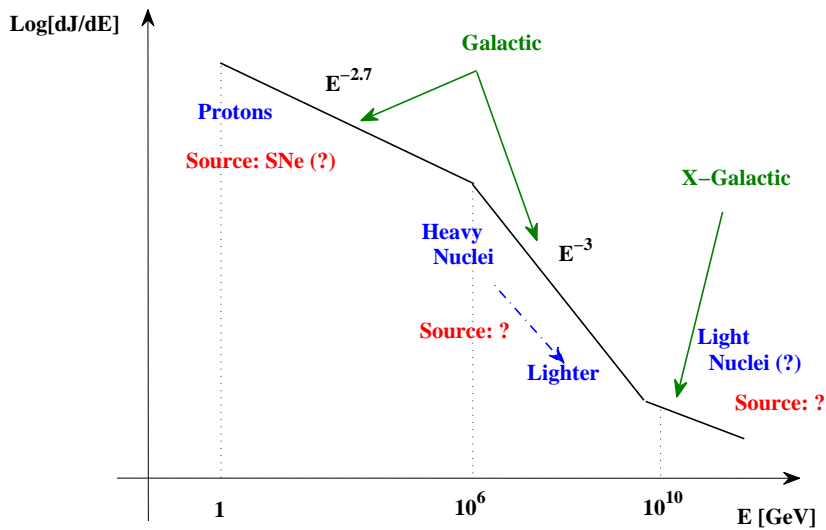}
%\vskip-2cm
\caption{A schematic description of the differential CR spectrum, $dJ/dE$, with some comments on what we know (or don't) about the composition and origin of the CRs.}
\label{fig:CR_pheno}
\end{figure}

The origin of CRs of all energies is still unknown (see \cite{Blandford87,Axford94,Nagano00} for reviews). The cosmic ray properties change qualitatively as a function of particle energy, as illustrated in Fig.~\ref{fig:CR_pheno}. The spectrum steepens around $\sim 5\times 10^{15}$~eV (the ``knee'') and flattens around $5\times 10^{18}$~eV (the ``ankle"). Below $\sim 10^{15}$~eV, the cosmic rays are thought to originate from Galactic supernovae. However, this hypothesis has not yet been confirmed (e.g. \cite{Butt_SNR_09} and references therein). The composition is dominated by protons at the lowest energies, and the fraction of heavy nuclei increases with energy. The proton fraction at $\sim 10^{15}$~eV is reduced to $\sim15\%$ \cite{Burnett90,Bernlohr98}. At yet higher energies, there
is evidence that the fraction of light nuclei increases, and that the cosmic-ray flux above $5\times 10^{18}$~eV is again dominated by protons \cite{Gaisser93,Bird94,Dawson98}. The composition change and the flattening of the spectrum around $10^{19}$~eV suggest that the flux above and below this energy is dominated by different sources. At energies of $E_{19}\equiv E/10^{19}{\rm eV}\sim1$ the Larmor radius of CRs in the Galactic magnetic field is
\begin{equation}
R_{L}=\frac{E}{ZeB}\approx 3 B_{-5.5}^{-1}E_{19}Z^{-1}{\rm kpc},
\end{equation}
where $B=10^{0.5}B_{-5.5}\mu$G is the value of the Galactic magnetic field and $Z$ is the CR charge. Since the Galactic magnetic field cannot confine protons above $10^{19}$~eV, it is believed that the nearly isotropic cosmic ray flux at $E>5\times 10^{18}$~eV originates from extra-Galactic (XG) sources. In what follows we focus on this XG component.

\subsection{Composition}
\label{ssec:primaries}

\begin{figure}
\hskip-.1cm\includegraphics[width=7.5cm]{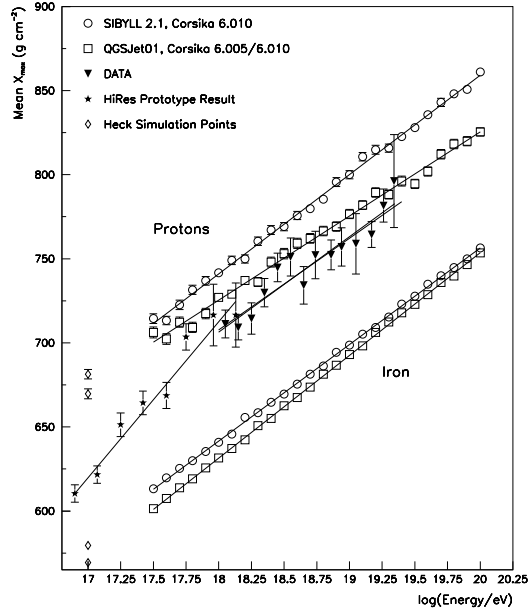}
\caption{Average depth of shower maximum as function of energy: Measurements by the HiRes detector compared to predictions for proton and iron primaries based on various model extrapolations of the $pp$ cross section. Adapted from \cite{HiRes_Composition_05}.}
\label{fig:comp_18}
\end{figure}

\begin{figure}
\hskip-.1cm\includegraphics[width=9cm]{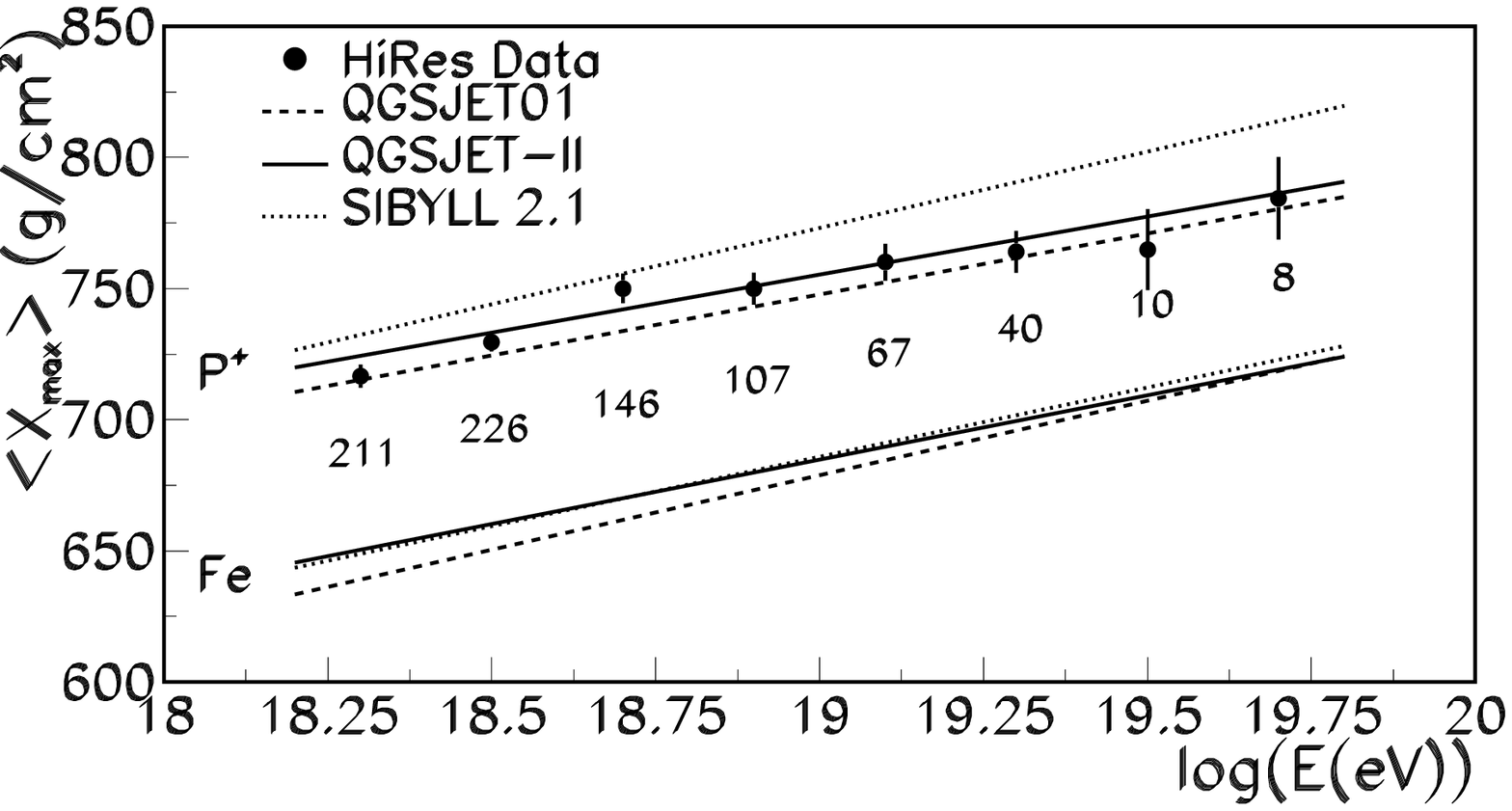}
\hskip-1cm\includegraphics[width=9cm]{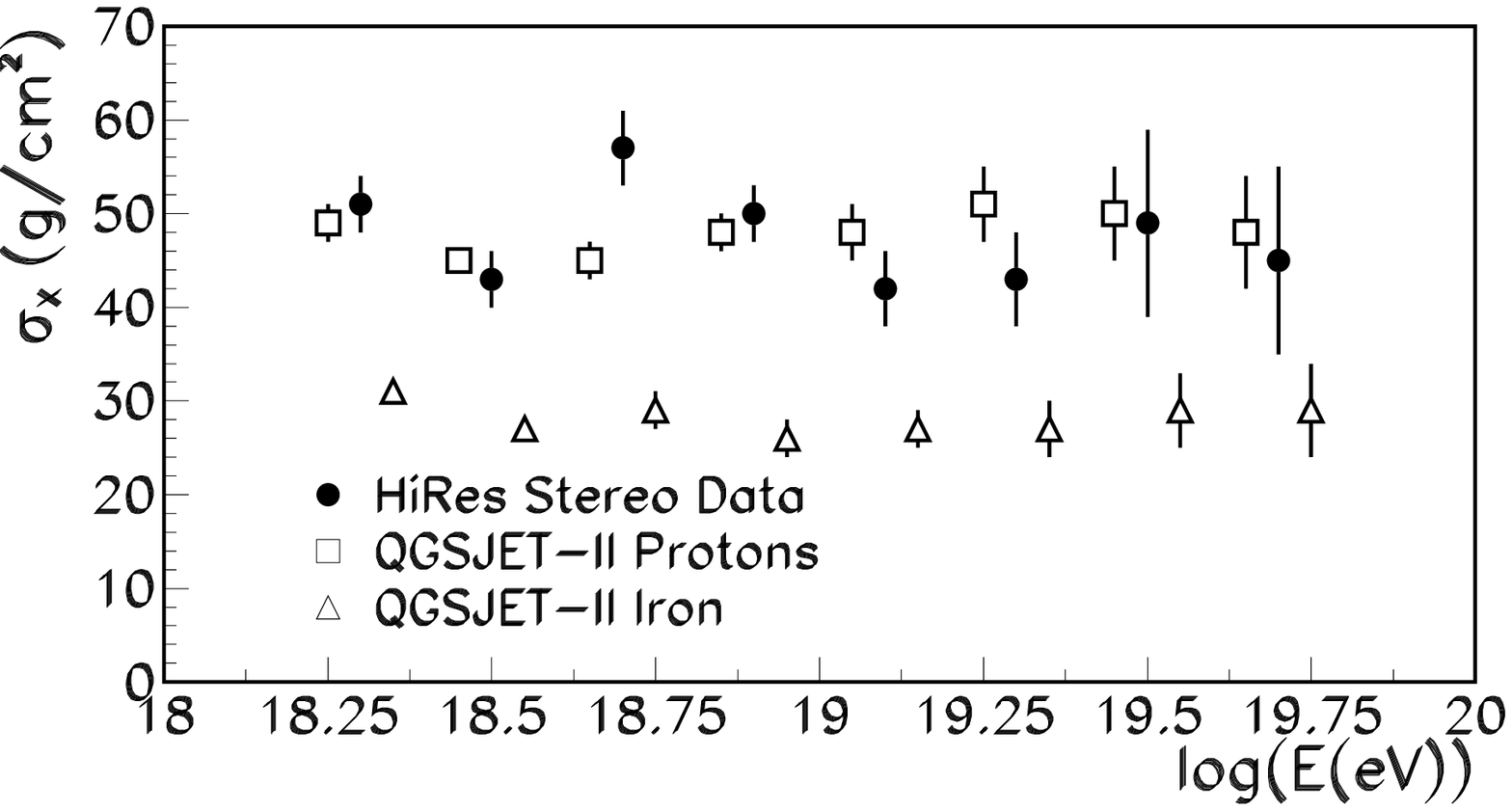}
\caption{Average and standard deviation of $X_{\rm max}$ at the highest energies: Measurements by the HiRes detector compared to model predictions. Adapted from \cite{HiRes_comp_2010}.}
\label{fig:comp_HiRes}
\end{figure}

\begin{figure}
\hskip-1cm\includegraphics[width=16cm]{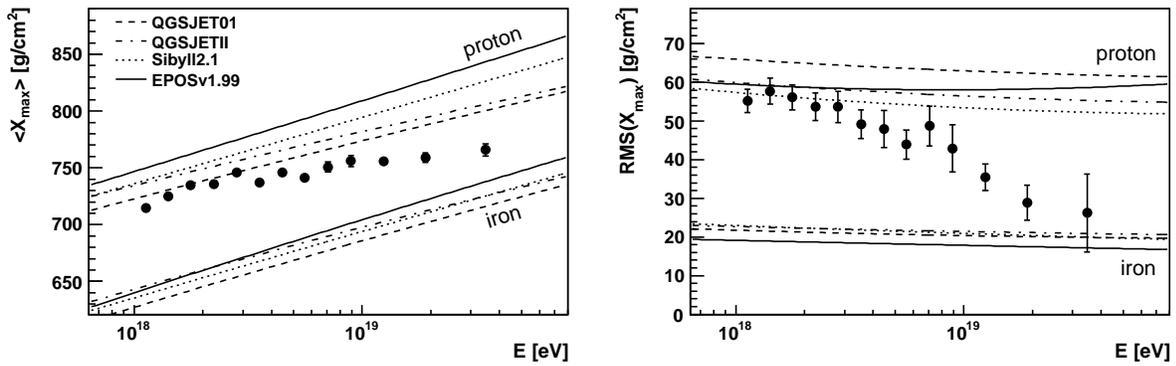}
\caption{Same as Fig.~\ref{fig:comp_HiRes}, for the PAO data.  Adapted from \cite{Auger_comp_2010}.}
\label{fig:comp_Auger}
\end{figure}

At low energy, $<10$~TeV, CR particles are detected by space or balloon born detectors, which provide a direct measurement of the primary CR composition. At higher energies, the flux is too low to be detectable by space/balloon born detectors, and CRs are detected indirectly through the ``air-showers", the large number of lower energy particles, they produce as they propagate and lose energy in the atmosphere. The low flux at the highest energies,
\begin{equation}\label{eq:J20}
    J(>10^{20}{\rm eV})\approx 1/100\,{\rm km^2 yr}\,2\pi\,{\rm sr},
\end{equation}
requires detectors with effective area of many 100's~km$^2$. The primary composition is constrained at high energies mainly by the average and variance of $X_{\rm max}$, the depth in the atmosphere at which the shower contains the largest number of high energy particles, obtained for showers of fixed energy (fluctuations in individual shower development are large, leading to fluctuations in the depth of maximum which are not small compared to the dependence on the primary mass).
$X_{max}$ is larger for higher energy particles. Since a high energy heavy nucleus behaves roughly as a group of independent lower energy nucleons, $X_{max}$ and its variance are larger at fixed energy for lighter nuclei.

Fig.~\ref{fig:comp_18} presents the main evidence for the transition to lighter nuclei at higher energy: $X_{\rm max}$ grows with energy faster than model predictions for fixed composition, becoming consistent with pure proton composition at $\sim10^{18}$~eV. At the highest observed energies, there is some discrepancy between the results reported by the HiRes observatory and by the Pierre Auger Observatory (PAO). While the HiRes observatory reports the average and variance of $X_{\rm max}$ to be consistent with a pure proton composition all the way up to $10^{19.7}$~eV (Fig.~\ref{fig:comp_HiRes}), the PAO reports $X_{\rm max}$ and $\sigma_X$ evolution which suggests a transition back to heavier nuclei at the highest energies (Fig.~\ref{fig:comp_Auger}).

The origin of this discrepancy is not yet understood. However, a few comments are in place. It was noted in \cite{WilkWlodarczyk10} that the analysis of the PAO data, presented in \cite{Auger_comp_2010}, is not self consistent: according to this analysis, $\sigma_X$ measured at the highest energy implies an Fe fraction $>90$\%, while the measured value of $<X_{\rm max}>$ implies an Fe fraction $<60$\%. This inconsistency may reflect some experimental problem, but may also reflect a modification of the hadronic interaction cross section which is not accounted for in the models used for shower calculations. It should be emphasized that the theoretical $X_{\rm max}$ calculations depend on extrapolation of hadronic models to energies well beyond those currently tested in accelerators. The theoretical and experimental uncertainties in the extrapolation of the $pp$ cross-section to center-of-mass energies
$\ge100\,$TeV are a possible source of biases in shower reconstruction (e.g. \cite{Ulrich09}). It is therefore difficult to draw a firm conclusion regarding primary composition at the highest energies based on current shower measurements.

\subsection{Generation rate \& spectrum}
\label{ssec:generation}

\begin{figure}
\includegraphics[width=8.5cm]{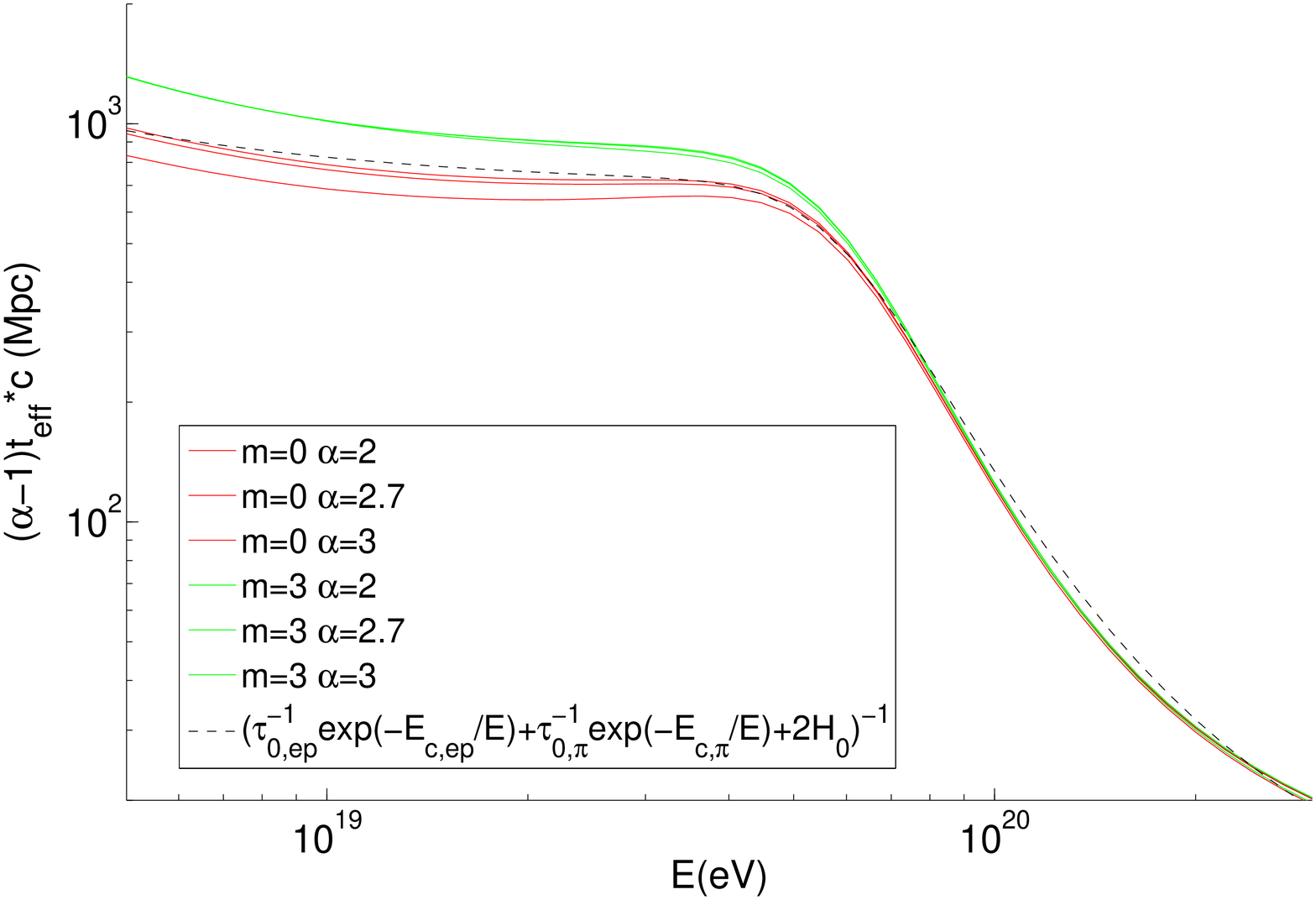}
\includegraphics[width=8.5cm]{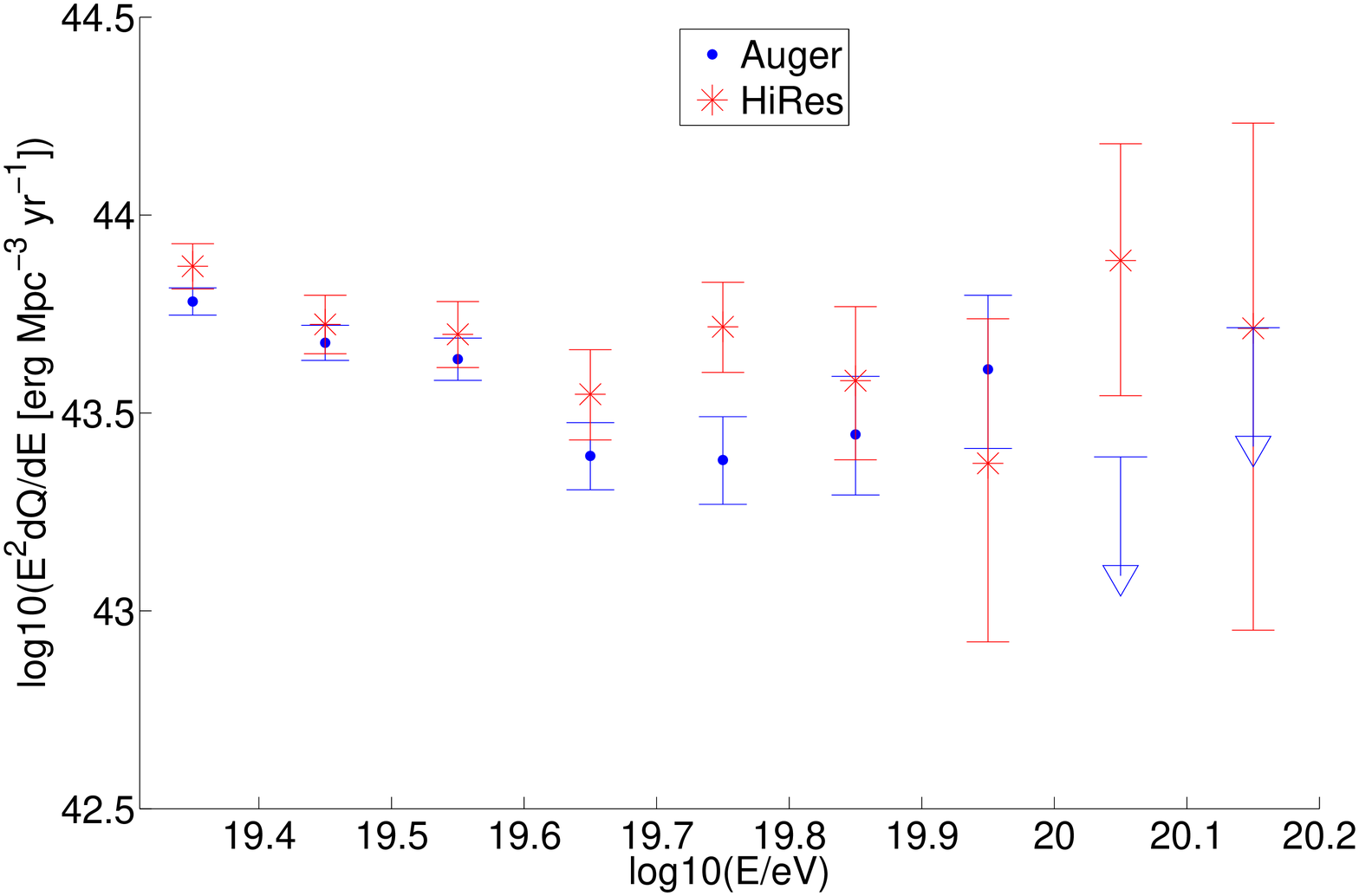}
\caption{{\it Left}: A comparison of direct numerical calculations of the effective CR life time (solid lines) with the analytic approximation of \cite{Katz_UHECR_09} using $\{E_{c,ep}=9.1\times10^{18}$~eV,~$\tau_{0,ep}=0.5\times10^9$~yr,~$E_{c,\pi}=3.5\times10^{20}$~eV,
$\tau_{0,\pi}=1.4\times10^7$~yr$\}$ (dashed line), for CR generation following $d\dot{n}/dE(E,z)\propto(1+z)^mE^{-\alpha}$. {\it Right}: The local ($z=0$) energy generation rate as measured by Auger \cite{Bluemer08} and Hires \cite{HiRes_GZK_08} assuming that the CRs are purely protons, for $\alpha-1=1$ (For different values of
$\alpha$, the spectrum should be multiplied by an energy independent factor $(\alpha-1)$; $Q\equiv\dot{n}$). Statistical and systematic errors in the experimental determination of event energies lead to $\sim 50\%$ uncertainty in the flux at the highest energies. The absolute energy scales of the Auger and Hires data where not altered in this figure. Adapted from \cite{Katz_UHECR_09}.}
\label{fig:generation_rate}
\end{figure}

Let us assume first that the UHECRs are protons of extra-Galactic origin. As they propagate, high-energy protons lose energy as a result of the cosmological redshift and as a result of production of pions and $e+e-$ pairs in interactions with cosmic microwave background (CMB) photons. The local intensity of UHECRs may be written as
\begin{equation}\label{eq:dJ/dE}
    \frac{dJ(E)}{dE}=\frac{c}{4\pi}\frac{d\dot{n}_0(E)}{dE}t_{\rm eff.}(E),
\end{equation}
where $d\dot{n}_0(E)/dE$ is the local ($z=0$) proton production rate (per unit volume and proton energy) and $t_{\rm eff.}$ is the effective energy loss time of the proton (this equation is, in fact, a definition of $t_{\rm eff.}$). The left panel of Fig.~\ref{fig:generation_rate} shows $t_{\rm eff.}$ for proton generation following $d\dot{n}/dE(E,z)\propto(1+z)^mE^{-\alpha}$. The rapid decrease in the effective life time, or propagation distance $ct_{\rm eff.}$, above $\sim6\times10^{19}$~eV, commonly termed the ``Greisen-Zatsepin-Kuzmin (GZK) suppression" \cite{Greisen,ZK}, is due to photo-production of pions by the interaction of protons with CMB photons (The proton threshold energy for pion production on $\sim10^{-3}$~eV CMB photons is $\sim10^{20}$~eV). Since proton propagation is limited at high energies to distances $\ll c/H_0$, e.g. to $\sim100$~Mpc at $10^{20}$~eV, the dependence of $t_{\rm eff.}$ on redshift evolution ($m$) is not strong.

Using Eq.~(\ref{eq:dJ/dE}) and the measured UHECR intensity, it is straightforward to infer the local production rate of UHECRs. The right panel of figure~\ref{fig:generation_rate} shows that the energy generation rate above $10^{19.5}$~eV is roughly constant per logarithmic CR energy interval, $\alpha\approx2$ and
\begin{equation}\label{eq:Q}
    E^2\frac{d\dot{n}_0(E)}{dE}\approx10^{43.5}{\rm erg/Mpc^3 yr}.
\end{equation}
In other words, the observed CR spectrum is consistent with a generation spectrum $d\dot{n}/dE\propto E^{-2}$ modified by the GZK suppression. Since both observations and models for particle acceleration in collisionless shocks, which are believed to be the main sources of high energy particles in many astrophysical systems, typically imply $\alpha\approx2$ (see \cite{Blandford87,Waxman_rel-plasma_rev06} for reviews of particle acceleration in non-relativistic and relativistic shocks respectively), this supports the validity of the assumption that UHECRs are protons produced by extra-Galactic objects.

The following point should, however, be made here. Heavy nuclei lose energy by interaction with CMB and IR photons, that leads to spallation. Since the effective life time of such nuclei is not very different from that of protons, the consistency of the observed spectrum with a model of extra-Galactic sources of protons with generation spectrum of $d\dot{n}/dE\propto E^{-2}$ can not be considered as a conclusive evidence for the UHECRs being protons.

One of the important open questions is at what energy the transition from Galactic to extra-Galactic (XG) sources takes place. A simple model with $E^2d\dot{n}_0(E)/{dE}=5\times10^{43}{\rm erg/ Mpc^{3}yr}$ and a transition from Galactic to XG sources at $10^{19}$~eV is consistent with observations \cite{Katz_UHECR_09}. In such a model, the Galactic flux is comparable to the XG one at $10^{19}$~eV, and negligible at $>10^{19.5}$~eV. Other models, however, have been proposed, in which the Galactic-XG transition occurs well below $10^{19}$~eV (e.g. \cite{Berezinsky08} and references therein). Such models are motivated mainly by the argument that they allow one to explain the $\sim5\times10^{18}$~eV spectral feature by pair production (in proton interactions with the CMB). The transition energy in such models is therefore well below $5\times10^{18}$~eV. As explained in \cite{Katz_UHECR_09}, a Galactic-XG transition at $\sim10^{18}$~eV requires fine tuning of the Galactic and XG contributions (to produce the smooth power-law observed), and is disfavored by the data: it requires that Auger systematically underestimates the energy of the events by 40\% (well above the stated uncertainty) and it requires $d\dot{n}_{p,\rm XG}/d\varepsilon\propto \varepsilon^{-2.7}$, which is inconsistent with the $>10^{19}$~eV data.

Finally, one notes that if the generation spectrum of XG CRs extends over many decades below $10^{19}$~eV, the total XG CR energy production rate, $Q_{\rm XG}^{z=0}$, might exceed significantly the UHECR production rate, $Q_{10^{19}\rm eV}^{z=0}\equiv (E^2d\dot{n}_0/dE)_{>10^{19}\rm eV}$, given by Eq.~(\ref{eq:Q}). For the $d\dot{n}/dE\propto1/E^2$ spectrum inferred from observations, the ``bolometric correction" will be $Q_{\rm XG}/Q_{10^{19} \rm eV}\approx\ln(10^{20}{\rm eV}/E_{\rm min})\sim10$, where $E_{\rm min}\ll10^{19}$~eV is the low energy to which the spectrum extends.

\subsection{Anisotropy: Source and composition clues}
\label{ssec:aniso}

\begin{figure}
\includegraphics[width=0.5\textwidth]{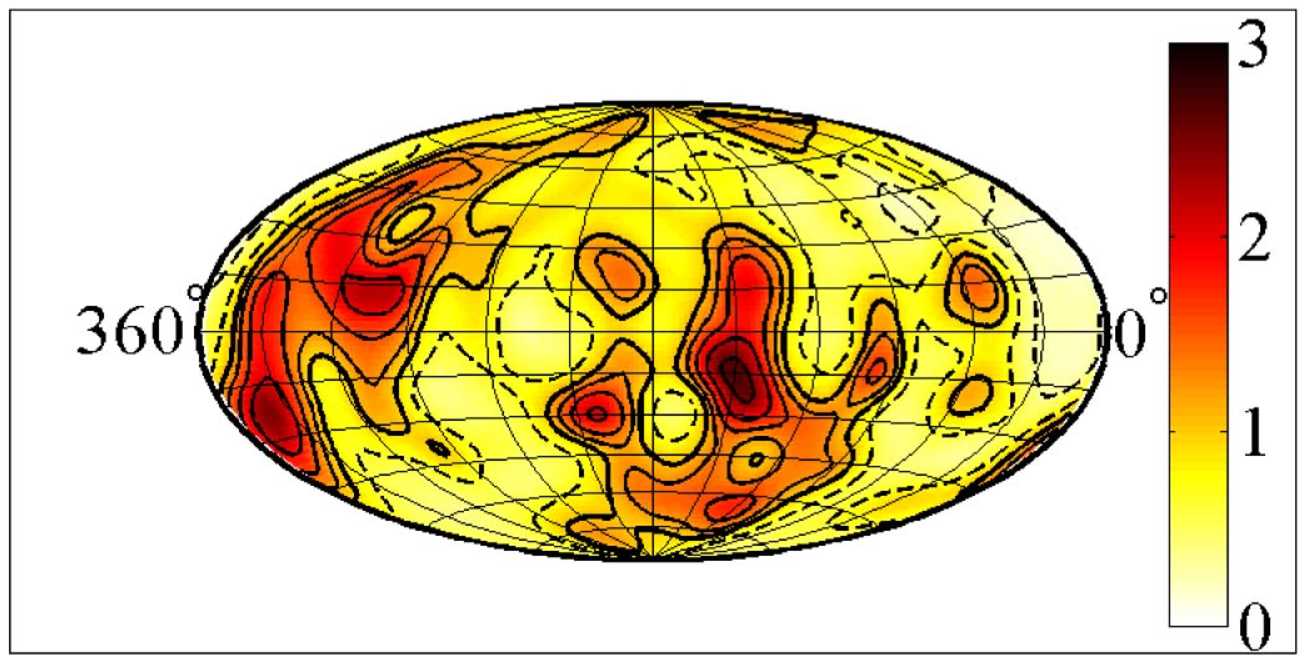}
\includegraphics[width=0.5\textwidth]{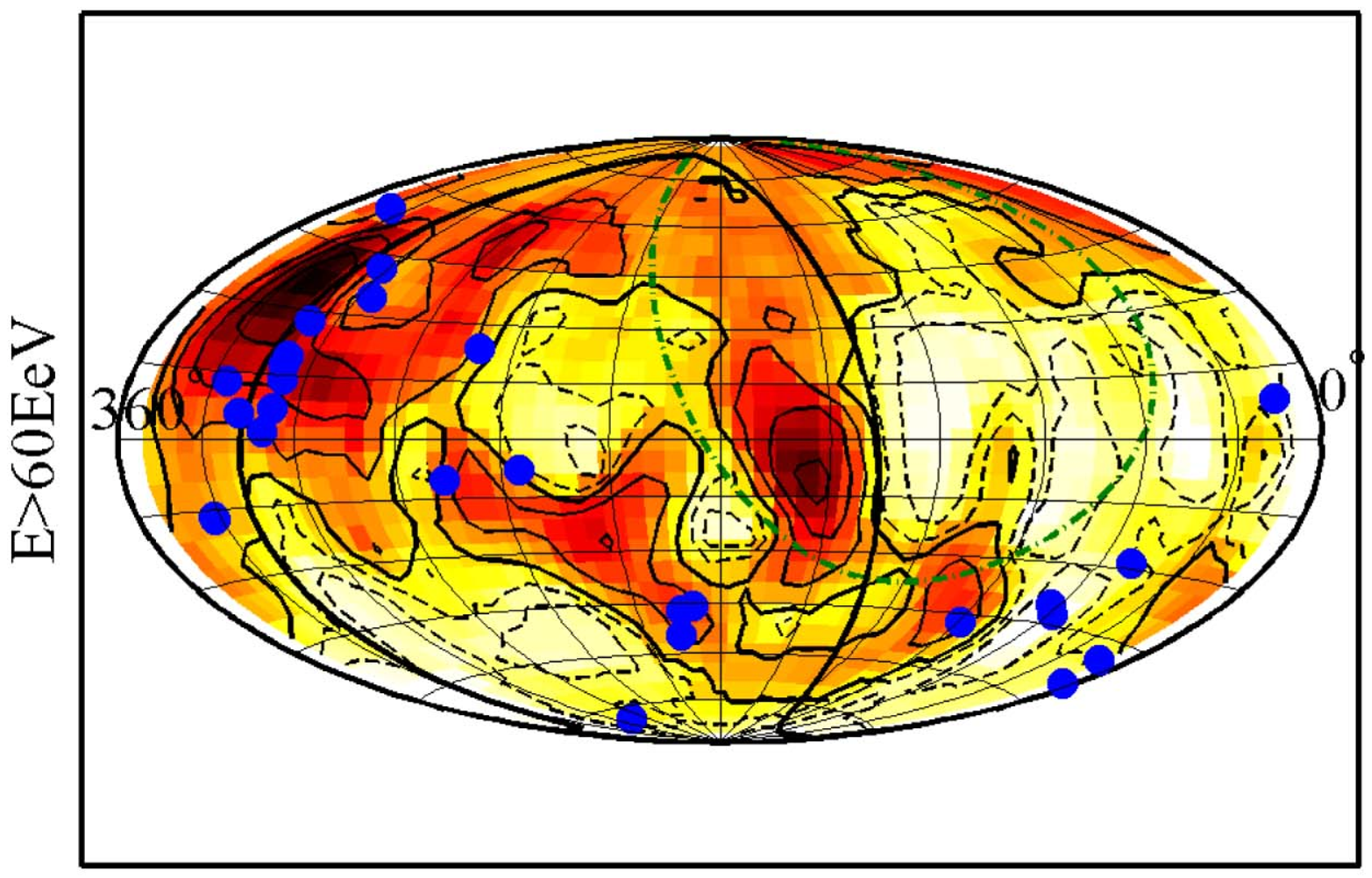}
\caption{{\it Top}: The integrated galaxy density out to a distance of $75$~Mpc, normalized to the mean integrated density. The contours are logarithmic, ranging from 0.5 to 4 with three contours per density doubling. Dashed curves represent under-density. {\it Bottom}: The positions of the 27 Auger events with energy exceeding $5.7\times10^{19}$~eV \cite{Auger_AGN_APh08}, overlaid on the UHECR intensity map, $J(\hat\Omega)$, predicted in a model in which the UHECR source distribution follows the galaxy density distribution (with a bias $b[\delta]=1+\delta$ for $\delta>0$, $b=0$ otherwise, where $\delta$ is the fractional galaxy over density). The coordinates are Galactic and $J$ is normalized to its all sky average. The contours denote $J/\bar{J}=(0.7,0.9,1,1.1,1.3,1.5)$, with dashed lines representing under-density. The thick solid line denotes the super-galactic plane. The dashed-dotted green line marks the boundary of Auger's coverage (corresponding to a zenith angle of $60^\circ$). Adapted from \cite{Kashti08}.}
    \label{fig:aniso}
\end{figure}

The propagation of UHECRs is limited at the highest energies to distances $\sim100$~Mpc. The galaxy distribution is not homogeneous over such a distance scale. Thus, if the distribution of UHECR sources is correlated with that of galaxies, one expects an anisotropy in the UHECR arrival direction distribution reflecting the inhomogeneity of the galaxy distribution \cite{WFP97}. Fig.~\ref{fig:aniso} shows the integrated galaxy density out to 75~Mpc and the predicted anisotropy of the UHECR intensity. Also shown are the (angular) positions of the 27 Auger events with energy exceeding $5.7\times10^{19}$~eV. The distribution of these events is inconsistent with isotropy at a $98\%$ confidence level for a source density $n_s=10^{-4}{\rm Mpc}^{-3}$, corresponding to the lowest allowed source density (see \S~\ref{ssec:density}), and at a $99\%$ confidence level for $n_s=10^{-2}{\rm Mpc}^{-3}$, corresponding to the density of galaxies. The angular distribution of CR arrival directions is consistent with a UHECR source distribution that follows the galaxy distribution (for detailed discussions see \cite{Kashti08,Takami09,Harari09,Koers09}). This provides some support to the association of the sources with known extra-Galactic astrophysical objects~\footnote{The evidence in the PAO data for a clustering of events in the region around Cen~A has triggered much discussion (see discussion and references in \cite{Lemoine09}). However, it is difficult to quantify the level of significance of the evidence for clustering, since it is based on an a posteriori analysis, as noted in~\cite{Auger_iso_09}. Moreover, one must keep in mind that Cen~A lies in front of one of the largest concentrations of matter in the local ($d\sim50\,$Mpc), Universe, $\{l=-51^\circ,b=19^\circ\}$, so that an excess of events from that direction does not necessarily imply that Cen~A is the source.}. The more recent PAO analysis (valid for $n_s\rightarrow\infty$) of a larger number of events, 58 above $5.5\times10^{19}$~eV, yields inconsistency with isotropy at a $99\%$ confidence level \cite{Auger_iso_09}.

%The dependence on the source number density may be understood by noting that the variance of the number of events $n$ in an angular bin, in which the number of sources is Poisson distributed with average $\bar{n}_{bs}$, and the number of events produced by each source is Poisson distributed with average $\lambda$, is $\bar{n^2}-\bar{n}^2=\bar{n}(1+\lambda)=\bar{n}(1+\bar{n}/\bar{n}_{bs})$, where $\bar{n}=\lambda\bar{n}_{bs}$. Thus, when the number of sources is not much larger than the number of events, the variance is larger than expected for a Poisson distribution of the event number. A comment should be made here regarding the more recent PAO analysis of a larger number of events, 58 above $5.5\times10^{19}$~eV, which yields inconsistency with isotropy at a $99\%$ confidence level \cite{Auger_iso_09}. The variance of $n$ in this analysis, in which the probability for having $n$ events (out of a total $N$ detected over all sky) in some angular bin is assumed to be binomial, $p(n)=[N!/n!(N-n)!]p_s^n(1-p_s)^{N-n}$ where $p_s\ll1$ is the probability for a single event to fall within the bin, is $\bar{n^2}-\bar{n}^2=\bar{n}(1-p_s)$ with $\bar{n}=Np_s$. Thus, the statistical significance of the inconsistency obtained by the PAO analysis is valid only for $\bar{n}/\bar{n}_{bs}\ll1$, i.e. for $n_s\gg10^{-5}{\rm Mpc}^{-3}$.

UHECRs may suffer significant deflections as they cross dense large scale structures, such as galaxy clusters and large scale galaxy filaments, in which the energy density of the plasma is large enough to support strong magnetic fields. Such deflections may distort the anisotropy pattern expected based on the galaxy distribution. An estimate of the expected deflection may be obtained assuming that all large scale structures support a magnetic field with energy density comprising a fraction $\epsilon_B$ of the plasma thermal energy density. The deflection expected in this case for a propagation distance $d$ is (see \cite{Kashti08,KoteraLemoine08} for a detailed derivation)
\begin{equation}
\label{eq:theta}
\theta\approx 0.3^\circ\frac{L}{1~\mbox{Mpc}}\left(\frac{f}{0.1}\frac{d}{100~\mbox{Mpc}}
\frac{\lambda}{10~\mbox{kpc}}\right)^{1/2}
\left(\frac{\epsilon_B}{0.01}\right)^{1/2}\left(\frac{E/Z}{10^{20}{\rm eV}}\right)^{-1}.
\end{equation}
Here, $Z$ is the particle charge, $f$ is the fraction of the volume filled by filaments of diameter $L$, and $\lambda$ is the field coherence length. The deflections are not expected therefore to distort significantly the anisotropy map.

The anisotropy signal provides also a test of the primary UHECR composition. If one records an anisotropy signal produced by heavy nuclei of charge $Z$ above an energy $E_{\rm thr}$, one should record an even stronger (possibly much stronger) anisotropy at energies $>E_{\rm thr}/Z$ due to the proton component that is expected to be associated with the sources of the heavy nuclei. This is due to the fact that particles of similar rigidity $E/Z$ propagate in a similar manner in the inter-galactic magnetic field and based on the plausible assumptions that (i) a source accelerating particles of charge $Z$ to energy $E$ will accelerate protons to energy $E/Z$, and (ii) there are at least as many protons accelerated as there are heavy nuclei. The anisotropy signal is expected to be stronger at lower energy since the signal increases as the number of particles produced by the source, $E^{-\alpha+1}$, while the background increases as the square-root of the number of all observed CRs, $\sim E^{-(2.7-1)/2}$ (see \cite{Lemoine09} for a detailed discussion). Thus, if the PAO $>5.7\times10^{19}$~eV anisotropy signal is real, the lack of detection of stronger anisotropy at lower energy disfavors a heavy nuclei composition at $\sim6\times10^{19}$~eV.

\subsection{Source density}
\label{ssec:density}

The arrival directions of the 27 PAO events and $\sim30$ HiRes events above $6\times10^{19}$~eV show no evidence for ``repeaters", i.e. multiple events that may be associated with a single source given the small deflection angles expected. The lack of repeaters implies that the number of sources contributing to the flux, $N_s$, should satisfy $N_s>N^2$, where $N$ is the number of events (for identical sources each producing on average $N/N_s$ events and $N^2/N_s\ll1$, the probability for repeaters is $\sim N^2/N_s$). This suggests that there should be more than $\sim10^{3.5}$ independent sources contributing to the (all sky) flux (note that HiReS and PAO observed the northern and southern hemispheres respectively). For protons, the effective propagation distance is $\sim200$~Mpc, see Fig.~\ref{fig:generation_rate}, implying a lower limit on the source density of
\begin{equation}\label{eq:n_s}
    n_s>10^{-4}{\rm Mpc^{-3}}
\end{equation}
(for a more detailed analysis see \cite{WFP97,Cuoco09}). For comparison, the density of galaxies is roughly $10^{-2}{\rm Mpc^{-3}}$.

\subsection{Source constraints: Minimum power and speed}
\label{ssec:power_gamma}

\begin{figure}
\includegraphics[width=0.6\textwidth]{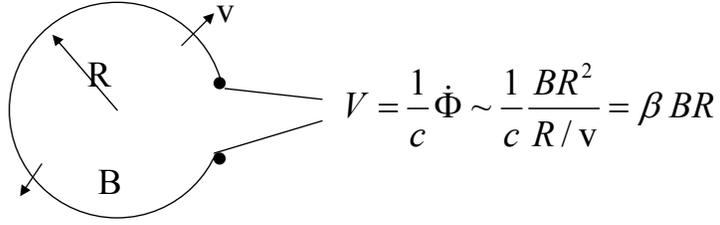}
\caption{Potential drop generated by an unsteady outflow of magnetized plasma.}
\label{fig:voltage}
\end{figure}

The essence of the challenge of accelerating particles to $>10^{19}$~eV can be understood using the following simple arguments (\cite{Waxman05}, for a more detailed derivation see \cite{Waxman95}). Consider an astrophysical source driving a flow of magnetized plasma, with characteristic magnetic field strength $B$ and velocity $v$. Imagine now a conducting wire encircling the source at radius $R$, as illustrated in Fig.~\ref{fig:voltage}. The potential generated by the moving plasma is given by the time derivative of the magnetic flux $\Phi$ and is therefore given by $V\approx\beta B R$ where $\beta=v/c$. A proton which is allowed to be accelerated by this potential drop would reach energy $E\sim\beta eB R$. The situation is somewhat more complicated in the case of a relativistic outflow, with $\Gamma\equiv(1-\beta^2)^{-1/2}\gg1$. In this case, the proton is allowed to be accelerated only over a fraction of the radius $R$, comparable to $R/\Gamma$. To see this, one must realize that as the plasma expands, its magnetic field decreases, so the time available for acceleration corresponds to the time of expansion from $R$ to, say, $2R$. In the observer frame this time is $R/c$, while in the plasma rest frame it is $R/\Gamma c$. Thus, a proton moving with the magnetized plasma can be accelerated over a transverse distance $\sim R/\Gamma$. This sets a lower limit to the product of the magnetic field and source size, which is required to allow acceleration to $E$, $BR>\Gamma E/e\beta$. This constraint also sets a lower limit to the rate $L$ at which energy should be generated by the source. The magnetic field carries with it an energy density $B^2/8\pi$, and the flow therefore carries with it an energy flux $>vB^2/8\pi$ (some energy is carried also as plasma kinetic energy), which implies $L>vR^2B^2$ and therefore
\begin{equation}\label{eq:L}
  L>\frac{\Gamma^2}{\beta}\left(\frac{E}{e}\right)^2c
  =10^{45.5}\frac{\Gamma^2}{\beta}\left(\frac{E}{10^{20}\rm eV}\right)^2{\rm erg/s}.
\end{equation}

Another constraint on the source results from the requirement that the acceleration is not suppressed by synchrotron emission of the accelerated particle. Let us consider a relativistic source. Denoting by $B'$ the magnetic field in the plasma rest frame, the acceleration time of a proton is $t'_{\rm acc}>E'/eB'c$ where $E'=E/\Gamma$ is the proton energy in the plasma frame. The synchrotron loss time, on the other hand, is given by $t'_{\rm syn}\approx (m_p/m_e)^2 (6\pi E'/\sigma_T c \gamma'^2 B'^2)$ where $\gamma'=E'/m_pc^2$. Requiring $ t'_{\rm acc}<t'_{\rm syn}$ sets an upper limit on $B'$ (which depends on $E$ and $\Gamma$). Requiring this upper limit to be larger than the lower limit derived in the previous paragraph, $B'R>E/e$, sets a lower limit to $\Gamma$ (which depends on $R$ and $E$). Relating the source radius $R$ to an observed variability time (of the radiation emitted by the source) through $R=2\Gamma^2c \delta t$, the lower limit is \cite{Waxman95}
\begin{equation}\label{eq:Gamma}
    \Gamma>10^2 \left(\frac{E}{10^{20}\rm eV}\right)^{3/4}\left(\frac{\delta t}{10\rm ms}\right)^{-1/4}.
\end{equation}
This implies that the sources must be relativistic, unless their characteristic variability time exceeds $\approx10^6$~s.

\subsection{Summary of source constraints}
\label{ssec:source_constraints}

The evidence for a transition to a light composition, consistent with protons, at few~$\times10^{18}$~eV (\S~\ref{ssec:primaries}), the consistency of the spectrum above $\sim10^{19}$~eV with a $d\dot{n}/dE\propto1/E^2$ generation spectrum modified by the GZK suppression (\S~\ref{ssec:generation}), and the hints for a light composition from the anisotropy signal (\S~\ref{ssec:aniso}), suggest that the UHECRs are protons produced by extra-Galactic sources. If this is indeed the case, the discussion of the preceding sections implies that their sources must satisfy several constraints:
\begin{itemize}
    \item The sources should produce protons with a local ($z=0$) rate and spectrum (averaged over space and time) $E^2d\dot{n}_0/dE\approx10^{43.5}{\rm erg/Mpc^3 yr}$;
    \item The density of sources (contributing to the flux at $\sim5\times10^{19}$~eV) should satisfy $n_s>10^{-4}{\rm Mpc^{-3}}$;
    \item The power output of the individual sources should satisfy $L>10^{45.5}\Gamma^2\beta^{-1}{\rm erg/s}$;
    \item The Lorentz factor of the flow driven by the source must satisfy $\Gamma>10^2 (\delta t/10{\rm ms})^{-1/4}$ where $\delta t$ is the characteristic source variability time.
\end{itemize}

No sources that satisfy the constraint $L>10^{46}{\rm erg/s}$ are known to lie within a $\sim100$~Mpc distance. One may argue, of course, that there are ``dark sources", i.e. sources that produce such power output (and UHECRs) but do not produce much radiation and are hence not known. One can not rule out the existence of such sources. On the other hand, we do not have direct evidence for their existence either. Putting aside such a caveat, the lack of known sources of sufficient luminosity suggests that the sources are transient. The transient duration $T$ must be shorter than the time delay between the arrival of photons and protons from the source. The protons are delayed due to magnetic field deflection by $\Delta t\sim \theta^2 d/c$, where $\theta$ is estimated in Eq.~(\ref{eq:theta}). This yields
\begin{equation}\label{eq:delay}
    \Delta t(E,d) \sim 10^4 \left(\frac{d}{100~\mbox{Mpc}}\right)^{2}\left(\frac{E}{10^{20}{\rm eV}}\right)^{-2} {\rm yr}.
\end{equation}
Due to the random energy loss of the protons during their propagation, and due to the possibility of multiple paths between source and observer, the arrival of protons of energy $E$ is delayed and spread over a similar time $\Delta t(E,d)$. For $T<\Delta t$, the effective number density of sources contributing to the flux at energy $E$ is $\sim\dot{n}_s\Delta t[E,d_{\rm eff}(E)]$, where $\dot{n}_s$ is the transient rate (per unit volume) and $d_{\rm eff}(E)\sim ct_{\rm eff}(E)$.

\subsection{``Suspects", Predictions}
\label{ssec:predictions}

Only two types of sources are known to satisfy the above minimum power requirement: active galactic nuclei (AGN) -- the brightest known steady sources, and gamma-ray bursts (GRBs) -- the brightest known transient sources\footnote{It was recognized early on (\cite{Hillas} and references therein) that while highly magnetized neutron stars may also satisfy the minimum power requirement, it is difficult to utilize the potential drop in their electro-magnetic winds for proton acceleration to ultra-high energy (see, however, \cite{Arons}).}. Both AGN (e.g. \cite{Rachen93}) and GRBs \cite{MU95,Vietri95,Waxman95} have therefore been suggested to be UHECR sources. The absence of AGN with $L>10^{46}~{\rm erg~s^{-1}}$ within the GZK horizon had motivated the suggestion \cite{Gruzinov} that UHECRs may be produced by a new, yet undetected, class of short duration AGN flares resulting from the tidal disruption of stars or accretion disk instabilities. The existence of tidal disruption flares is likely. However, they are yet to be detected and whether their properties are consistent with the constraints derived above is yet to be determined (see also \cite{WL09}).

Let us consider then the GRB transients. First, consider the minimum power and minimum speed constraints that should be satisfied by individual sources: Eqs.~(\ref{eq:L}) and~(\ref{eq:Gamma}). For GRBs, the (luminosity function averaged) peak luminosity is $L_\gamma\approx10^{52}{\rm erg/s}$ (\cite{Guetta05,Wanderman10}, note that \cite{Guetta05} gives $L_{\rm 50-300~keV}$ which is $\approx0.1$ of $L_{\rm 0.1-10~MeV}$ given in \cite{Wanderman10}), and typical values of $\Gamma$ and $\delta t$ are $\Gamma\simeq10^{2.5}$ and $\delta t\sim10$~ms \cite{Meszaros02,Waxman03,Piran04,Meszaros06}. Thus, both constraints are satisfied. It is worth noting that $\Gamma>10^2$ is inferred for GRBs based on the photon spectrum (in order to avoid large pair production optical depth), i.e. based on arguments which are different than those leading to the $\Gamma>10^2$ constraint of Eq.~(\ref{eq:Gamma}).

Next, let us consider the global constraints on the rate, Eq.~(\ref{eq:n_s}), and average energy production rate, Eq.~(\ref{eq:Q}), of the sources. The local, $z=0$, GRB rate is $\dot{n}_s^{z=0}\sim10^{-9}{\rm Mpc^{-3}yr^{-1}}$ (assuming $\dot{n}_s$ evolves rapidly with redshift, following the star formation rate, i.e $\dot{n}_s^{z=0}\ll \dot{n}_s^{z=1.5}$ \cite{Guetta05,Wanderman10}), implying, using Eq.~(\ref{eq:delay}), $n_s(E)\sim\dot{n}_s^{z=0}\Delta t[E,d_{\rm eff}(E)]\sim10^{-4}(d_{\rm eff}/200{\rm Mpc})^2(E/0.5\times10^{20}{\rm eV})^{-2}{\rm Mpc}^{-3}$, consistent with Eq.~(\ref{eq:n_s}). The local, $z=0$, GRB energy production rate in $\sim1$~MeV photons is given by $\dot{n}_s^{z=0} L_\gamma\Delta t$, where $\Delta t$ is the effective duration (the average ratio of the fluence to the peak luminosity) corrected for redshift (the observed duration is $1+z$ larger than the duration at the source), $\Delta t\approx10{\rm s}/(1+z)\sim4$~s (using $z=1.5$ as a characteristic redshift). This yields $E_\gamma\equiv L_\gamma\Delta t\approx10^{52.5}{\rm erg}$ and $Q_{\rm MeV, GRB}^{z=0}\equiv\dot{n}_s^{z=0} E_\gamma\approx10^{43.5}{\rm erg/Mpc^3 yr}$, similar to the required UHECR energy production rate given in Eq.~(\ref{eq:Q}), $Q_{10^{19}\rm eV}^{z=0}\equiv (E^2 d\dot{n}_0/dE)_{>10^{19}\rm eV}\approx10^{43.5}{\rm erg/Mpc^3 yr}$ (for a more detailed discussion see \cite{W04_GRB_CR,Le07,W10_GRB_CR}; for additional energy production by ``low-luminosity GRBs" and ``heavy baryon loading GRBs" see \cite{MuraseIoka08} and \cite{Wick04} respectively, and references therein).

As noted at the end of \S~\ref{ssec:generation}, if the generation spectrum of XG CRs extends over many decades below $10^{19}$~eV, the total XG CR energy production rate, $Q_{\rm XG}^{z=0}$, may exceed significantly the UHECR production rate, $Q_{\rm XG}/Q_{10^{19}\rm eV}\sim10$. Estimating the ratio of $Q_{\rm XG}^{z=0}$ to the total photon energy production by GRBs, $Q_{\gamma,\rm GRB}^{z=0}$, as
$Q_{\rm XG}^{z=0}/Q_{\gamma,\rm GRB}^{z=0}=Q_{\rm XG}^{z=0}/Q_{\rm MeV, GRB}^{z=0}\sim10$ is, however, quite uncertain. This is due to uncertainties in the redshift evolution of the GRB rate and luminosity function, in the ``bolometric correction" for the CR production rate, and in the bolometric correction, $Q_{\gamma,\rm GRB}^{z=0}/Q_{\rm MeV, GRB}^{z=0}>1$, that should also be applied to the photons.

If GRBs are the sources of UHECRs, then some interesting predictions can be made regarding the spectrum and angular distribution of events at the highest energies \cite{Miralda96}. Due to the rapid decrease of $d_{\rm eff}$ with energy, the total number of sources contributing to the flux, $\sim (4\pi/3)d_{\rm eff}^3\dot{n}_s\Delta t[E,d_{\rm eff}(E)]$ drops rapidly with energy. This implies that, for $\dot{n}_s\sim10^{-9}{\rm Mpc^{-3}yr^{-1}}$ and adopting the estimate of Eq.~(\ref{eq:theta}) for the deflection angle, only a few sources contribute to the flux above $\sim3\times10^{20}$~eV. Moreover, the spectrum of these sources should be rather narrow, $\Delta E/E\sim1$, since the energy dependent time delay $\Delta t(E,d)$ implies that higher (lower) energy particles arrived (will arrive) in the past (future). Testing this prediction, which requires a large number of events detected above $\sim3\times10^{20}$~eV,  may require large exposure, exceeding even that of PAO, which may be provided by space born detectors \cite{Space_CR_08,JEM-EUSO09}.

\section{High energy neutrino astronomy}
\label{sec:nu}

UHECR sources are likely to be sources of high energy neutrinos. The interaction of high energy protons (nucleons) with radiation or gas, either at or far from the source, leads to production of charged pions, via $p\gamma$ and $pp(n)$ interactions, which decay to produce neutrinos (e.g. $p+\gamma\rightarrow n+\pi^+$, $\pi^+\rightarrow\mu^++\nu_\mu\rightarrow e^++\nu_\mu+\bar{\nu}_\mu+\nu_e$). In \S~\ref{ssec:bound} we estimate the minimum detector size, which is required to detect such neutrinos.  In \S~\ref{ssec:GZK} we comment on the importance of the detection of ``GZK neutrinos". The prospects for detection of GRB neutrinos, and the possible implications of such detection for the study of GRBs, are discussed in \S~\ref{ssec:grb_nu}. Prospects for the study of fundamental neutrino properties using high energy GRB neutrinos are discussed in \S~\ref{ssec:nu_phys}. For most of the discussion of this section, we adopt the assumption that UHECRs are protons.

\subsection{Neutrino flux upper bound, Detector size, Detectors' status}
\label{ssec:bound}

\begin{figure}
\includegraphics[width=0.5\textwidth]{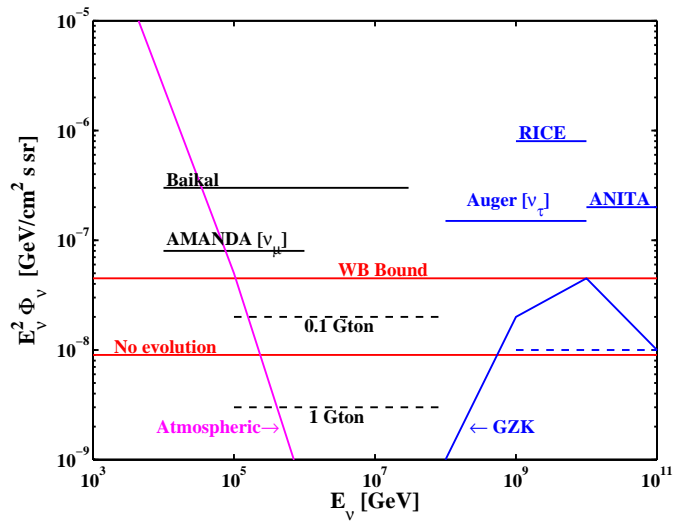}
\caption{The upper bound imposed by UHECR observations on the extra-Galactic high energy muon neutrino ($\nu_\mu+\bar\nu_\mu$) intensity \cite{WBbound1,WBbound2} (lower-curve: no evolution of the energy production rate, upper curve: assuming evolution following star formation rate), compared with the atmospheric muon-neutrino background and with several experimental upper bounds (solid lines). The theoretical bound does not include the effect of neutrino oscillations. Such oscillations are expected to change the $\nu_e:\nu_\mu:\nu_\tau$ flavor ratio from $1:2:0$ to $1:1:1$ (e.g. \cite{Learned95}), leading to an upper bound which is $\approx1/2$ that shown in the figure for each flavor. Shown are the muon and all flavor upper bounds of the optical Cerenkov observatories AMANDA \cite{Achterberg07,Achterberg08} and BAIKAL \cite{Avrorin09}, the all flavor upper bounds of the coherent Cerenkov radio detectors RICE  \cite{Kravchenko} and ANITA \cite{Gorham09}, and the $\nu_\tau$ upper bound of the PAO \cite{Abraham09}. The curve labelled ``GZK" shows the muon neutrino intensity (not corrected for oscillations) expected from UHECR proton interactions with micro-wave background photons \cite{gzk_nu}. Black dashed curves show the expected sensitivity (for few years operation) of 0.1~Gton (ANTARES, http://antares.in2p3.fr/) and 1~Gton (IceCube, http://icecube.wisc.edu/; Km3Net, http://www.km3net.org/home.php) optical Cerenkov detectors. The blue dashed curve is the expected sensitivity of detectors of few 100~Gton (few 100~km$^3$) effective mass (volume), that may be achieved with proposed radio detectors \cite{Barwick07,Landsman08,Allison09,Barwick09} or with proposed (optical) extensions of IceCube \cite{HalzenHooper04}. For a detailed discussion of the current experimental status see \cite{Hoffman_nu_tel_09,Anchordoqui09}.}
\label{fig:bound}
\end{figure}

The energy production rate, Eq.~(\ref{eq:Q}), sets an upper bound to the neutrino intensity produced by sources which, as GRBs and AGN jets, are for high-energy nucleons optically thin to $p\gamma$ and $pp(n)$ interactions. For sources of this type, the energy generation rate of neutrinos can not exceed the energy generation rate implied by assuming that all the energy injected as high-energy protons is converted to pions (via $p\gamma$ and $pp(n)$ interactions). Using Eq.~(\ref{eq:Q}), the resulting upper bound ($\nu_\mu+\bar\nu_\mu$, neglecting mixing) is \cite{WBbound1,WBbound2}
\begin{equation}
E_\nu^2 \Phi_\nu < \frac{1}{4}\xi_Zt_H{c\over4\pi}E^2{d\dot n_0\over
dE} \approx10^{-8}\xi_Z\left(\frac{E^2d\dot n_0/dE}{10^{44}{\rm erg/Mpc^3yr}}\right)\,{\rm GeV\,cm}^{-2}{\rm s}^{-1}{\rm sr}^{-1}.
\label{eq:WB}
\end{equation}
Here $t_H$ is the Hubble time and the $1/4$ factor is due to the fact that charged and neutral pions (which decay to photons) are produced with similar probability, and that muon neutrinos carry roughly half the energy of the decaying pion. In the derivation of Eq.~(\ref{eq:WB}) we have neglected the
redshift energy loss of neutrinos produced at cosmic time $t<t_H$,
and implicitly assumed that the cosmic-ray generation rate per
unit (comoving) volume is independent of cosmic time. The quantity
$\xi_Z$ in Eq.~\ref{eq:WB} has been introduced to describe
corrections due to redshift evolution and energy loss.

The upper bound is compared in Fig.~\ref{fig:bound} with the current experimental limits and with the expected sensitivity of planned neutrino telescopes. The figure indicates that km-scale (i.e. giga-ton-scale) neutrino telescopes are needed to detect the expected extra-Galactic flux in the energy range of $\sim1$~TeV to $\sim1$~PeV, and that much larger effective volume is required to detect the flux at higher energy. The Baikal, AMANDA, and ANTARES optical Cerenkov telescopes have proven that the construction of km-scale neutrino detectors is feasible, and the IceCube detector, the construction of which is well underway, is expected to reach its designed target effective mass of $\sim1$~Gton in 2011.

\subsection{GZK neutrinos}
\label{ssec:GZK}

As discussed in \S~\ref{ssec:generation}, protons of energy exceeding the threshold for pion production in interaction with CMB photon, $\sim5\times10^{19}$~eV, lose most of their energy over a time short compared to the age of the universe. If UHECRs are indeed protons of extra-Galactic origin, their energy loss should produce a neutrino intensity similar to the upper bound given by Eq.~(\ref{eq:WB}). Since most of the pions are produced in interactions with photons of energy corresponding to the $\Delta$-resonance, each of the resulting neutrinos carry approximately 5\% of the proton energy. The neutrino background is therefore close to the bound above $\sim5\times10^{18}$~eV, where neutrinos are produced by $\sim10^{20}$~eV protons. The intensity at lower energies is lower, since protons of lower energy do not lose all their energy over the age of the universe (The GZK intensity in figure~\ref{fig:bound} decreases at the highest energies since it was assumed that the maximum energy of protons produced by UHECR sources is $10^{21}$~eV). The results of detailed calculations of the expected GZK neutrino intensity \cite{ESS01} are in agreement with the qualitative analysis presented above.

The detection of GZK neutrinos will be a milestone in neutrino astronomy. Most important, it will allow one to test the hypothesis that the UHECRs are protons (possibly somewhat heavier nuclei) of extra-Galactic origin (e.g. \cite{MuraseBeacom10} and references therein). Moreover, measurements of the flux and spectrum would constrain the redshift evolution of the sources. Finally, detection of ultra-high energy neutrinos may allow one to test for modifications of the neutrino interaction cross section due to new physics effects at high (100~TeV) energies \cite{KusenkoWeiler02,Anchordoqui_nu_cs06,Barwick09}.

\subsection{Neutrinos from GRBs}
\label{ssec:grb_nu}

GRB gamma-rays are believed to be produced within a relativistic expanding wind, a so called ``fireball", driven by rapid mass accretion onto a newly formed stellar-mass black hole. It is commonly assumed that electrons are accelerated to high energy in collisionless shocks taking place within the expanding wind, and that synchrotron emission from these shock accelerated electrons produces the observed $\gamma$-rays (see \cite{Meszaros02,Waxman03,Piran04,Meszaros06} for reviews). If protons are present in the wind, as assumed in the fireball model, they would also be accelerated to high energy in the region were electrons are accelerated. If protons are indeed accelerated, then high energy neutrino emission is also expected.

\subsubsection{100 TeV fireball neutrinos}
\label{ssec:fireball_nu}

Protons accelerated in the region where MeV gamma-rays are produced will interact with these photons to produce pions provided that their energy exceeds the threshold for pion production,
\begin{equation}
E_\gamma \,E \approx 0.2 \Gamma^2\, {\rm GeV^2} \,.
\label{eq:keyrelation}
\end{equation}
Here, $E_\gamma$ is the observed photon energy. The $\Gamma^2$ factor appears since the protons and photon energies in the plasma rest frame (where the particle distributions are roughly isotropic) are smaller than the observed energy by the Lorentz factor $\Gamma$ of the outflow. For $\Gamma\approx10^{2.5}$ and $E_\gamma=1$~MeV, proton energies $\sim 10^{16}$~eV are required to produce pions. Since neutrinos produced by pion decay typically carry $5\%$ of the proton energy, production of $\sim 10^{14}$~eV neutrinos is expected \cite{WnB97}.

\begin{figure}
\includegraphics[width=0.4\textwidth]{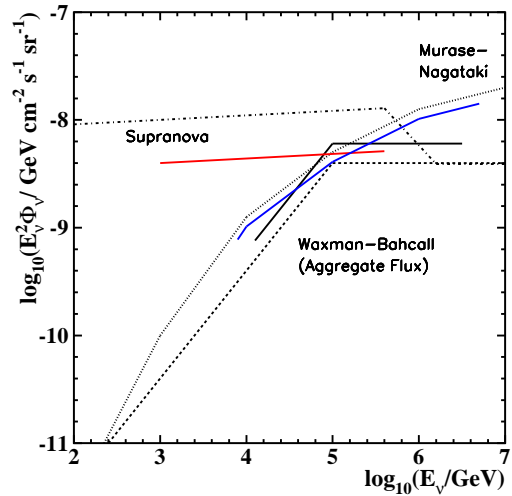}
\caption{AMANDA flux upper limits (solid lines, 90\% confidence) for muon neutrino energy
spectra predicted by the models of \cite{WnB97,MuraseNagataki06} for the $\sim100$~TeV internal shock fireball neutrinos (\S~\ref{ssec:fireball_nu}), and for the muon neutrino energy spectrum predicted by \cite{Razzaque03} for the precursor supernova (``supranova") model (\S~\ref{ssec:TeV}). The upper bounds are compared with the fluxes predicted by the models (\cite{WnB97,W03_XG_nu_review}- thick dotted line, \cite{MuraseNagataki06}- thin dotted line, \cite{Razzaque03}- dot-dashed line). Adapted from \cite{Achterberg08_Amanada_GRB}.}
\label{fig:GRB_nu}
\end{figure}

The fraction of energy lost by protons to pions, $f_\pi$, is $f_\pi\approx0.2$ \cite{WnB97,GSW01}. Assuming that GRBs generate the observed UHECRs, the expected GRB muon and anti-muon neutrino flux may be estimated using Eq.~(\ref{eq:WB}) \cite{WnB97,WBbound1},
\begin{equation}
E_\nu^2\Phi_{\nu}\approx 10^{-8}{f_\pi\over0.2}\left(\frac{E^2d\dot n_0/dE}{10^{44}{\rm erg/Mpc^3yr}}\right){\rm GeV\,cm}^{-2}{\rm s}^{-1}{\rm
sr}^{-1}. \label{eq:JGRB}
\end{equation}
This neutrino spectrum extends to $\sim10^{16}$~eV, and is suppressed at higher energy due to energy loss of pions and muons \cite{WnB97,RnM98,WBbound1} (for the contribution of Kaon decay at high energy see \cite{Asano06}). Eq.~(\ref{eq:JGRB}) implies a detection rate of $\sim10$ neutrino-induced muon events per year (over $4\pi$~sr) in a 1~Gton (1~cubic-km) detector \cite{WnB97,Alvarez00,Guetta04,BeckerHalzen06,MuraseNagataki06}. The upper limit on the GRB neutrino emission provided by the AMANDA ($\sim0.05$~Gton) detector approaches the flux predicted by Eq.~(\ref{eq:JGRB}), see Fig.~\ref{fig:GRB_nu}, and the 1~Gton IceCube detector, which will be completed at the beginning of 2011, will reach a sensitivity that may allow one to test this model's predictions \cite{IceCube_GRB_nu}.

Since GRB neutrino events are correlated both in time and in direction with gamma-rays, their detection is practically background free. The main background is due to atmospheric neutrinos, which produce neutrino-induced muons, travelling in a direction lying within a cone of opening angle $\Delta\theta$ around some direction, at a rate
\begin{equation}\label{eq:atmo}
    J_{\nu\rightarrow\mu}^A\simeq4\times10^{-3}\left(\frac{\Delta\theta}{0.5^o}\right)^2
\left(\frac{E}{100\rm TeV}\right)^{-\beta}\,{\rm km^{-2}yr^{-1}},
\end{equation}
with $\beta=1.7$ for $E<100$~TeV and $\beta=2.5$ for $E>100$~TeV. At high energies, the neutrino induced muon propagates at nearly the same direction as the incoming neutrino, and km-scale neutrino telescopes will be able to determine the incoming neutrino direction to better than $\sim0.5^o$. For a known source direction, therefore, the neutrino search is practically background free.

\subsubsection{TeV neutrinos}
\label{ssec:TeV}

The 100~TeV neutrinos discussed in the previous sub-section are produced in the same region where GRB $\gamma$-rays are produced and should therefore accompany the 10 to 100~s $\gamma$-ray emission phase (note, however, that it was pointed out in \cite{MuraseNagataki06b} that if the late, $\sim10^4$~s, X-ray/UV flares are produced by late internal shocks within the fireball, the emission of 100~TeV neutrinos may be extended to accompany these flares). Their production is a generic prediction of the fireball model: it is a direct consequence of the assumptions that energy is carried from the underlying engine as kinetic energy of protons and that $\gamma$-rays are produced by synchrotron emission of shock accelerated particles. Neutrinos may be produced also in other stages of fireball evolution, at energies different than 100~TeV. The production of these neutrinos is dependent on additional model assumptions. We discuss below some examples of $\sim1$~TeV neutrino emission predictions, that depend on the properties of the GRB progenitor. For a discussion of $\sim10^{18}$~eV neutrino emission during the afterglow phase see \cite{Vietri98,WnB00,Dai01,Murase07} and the reviews \cite{Meszaros02,W03_XG_nu_review,Meszaros06}.

The most widely discussed progenitor scenarios for long-duration GRBs involve core collapse of massive stars. In these ``collapsar" models, a relativistic jet breaks through the stellar envelope to produce a GRB. For extended or slowly rotating stars, the jet may be unable to break through the envelope. Both penetrating (GRB producing) and ``choked" jets can produce a burst of $\sim10$~TeV neutrinos by interaction of accelerated protons with jet photons, while the jet propagates in the envelope \cite{MnW01,Razzaque04,AndoBeacom05} (it was pointed out in \cite{AndoBeacom05} that neutrino production by kaon decay may dominate over the pion decay contribution, extending the neutrino spectrum to $\sim20$~TeV). The estimated event rates may exceed $\sim10^2$ events per yr in a km-scale detector, depending on the ratio of non-visible to visible fireballs. A clear detection of non-visible GRBs with neutrinos may be difficult due to the low energy resolution for muon-neutrino events, unless the associated supernova photons are detected.

In the two-step ``supranova" model, interaction of the GRB blast wave with the supernova shell can lead to detectable neutrino emission, either through nuclear collisions with the dense supernova shell or through interaction with the intense supernova and backscattered radiation field \cite{Dermer03,Guetta03,Razzaque03}. As indicated by Fig.~\ref{fig:GRB_nu}, the upper limits provided by AMANADA on the muon neutrino flux suggest that ``supranova"s do not accompany most GRBs.

\subsection{Neutrino physics prospects}
\label{ssec:nu_phys}

In addition to testing the GRB model for UHECR production and to providing a new handle on the physics of GRB sources, detection of high energy GRB neutrinos may provide information on fundamental neutrino properties \cite{WnB97}.

Detection of neutrinos from GRBs could be used to test the simultaneity of neutrino and photon arrival to an accuracy of $\sim1$~s. It is important to emphasize here that since the background level of
neutrino telescopes is very low, see Eq.~(\ref{eq:atmo}), the detection of a single neutrino from the direction of a GRB on a time sale of months after the burst would imply an association of the neutrino with the burst and will therefore establish a time of flight delay measurement. Such a measurement will allow one to test for violations of Lorentz invariance (as expected due to quantum gravity effects) \cite{WnB97,Amelino98,Coleman99,Jacob07}), and to test the weak equivalence principle, according to which photons and neutrinos should suffer the same time delay as they pass through a gravitational potential. With $1{\rm\ s}$ accuracy, a burst at $1{\rm\ Gpc}$ would reveal a fractional difference in (photon and neutrino) speed of $10^{-17}$, and a fractional difference in gravitational time delay of order $10^{-6}$ (considering the Galactic potential alone). Previous applications of these ideas to supernova 1987A (see \cite{Bahcall89} for review), yielded much weaker upper limits: of order $10^{-8}$ and $10^{-2}$ respectively. Note that at the high neutrino energies under discussion deviations of the propagation speed from that of light due to the finite mass of the neutrino lead to negligible time delay even from propagation over cosmological distances (less than $\sim10^{-10}$~s at 100~TeV).

High energy neutrinos are expected to be produced in GRBs by the decay of charged pions, which lead to the production of neutrinos with flavor ratio $\Phi_{\nu_e}:\Phi_{\nu_\mu}:\Phi_{\nu_\tau}=1:2:0$ (here $\Phi_{\nu_l}$ stands for the combined flux of $\nu_l$ and $\bar\nu_l$). Neutrino oscillations then lead to an observed flux ratio on Earth of
$\Phi_{\nu_e}:\Phi_{\nu_\mu}:\Phi_{\nu_\tau}=1:1:1$ \cite{Learned95,Athar06} (see, however \cite{Kashti05}). Up-going $\tau$'s, rather than $\mu$'s, would be a distinctive signature of such oscillations. It has furthermore been been pointed out that flavor measurements of astrophysical neutrinos may help determining the mixing parameters and mass hierarchy \cite{Winter06}, and may possibly enable one to probe new physics \cite{Learned95,Athar00}.

\section{Outlook: Open questions \& Multi-messenger astronomy}
\label{sec:multi-messenger}

\begin{figure}
\includegraphics[width=0.4\textwidth]{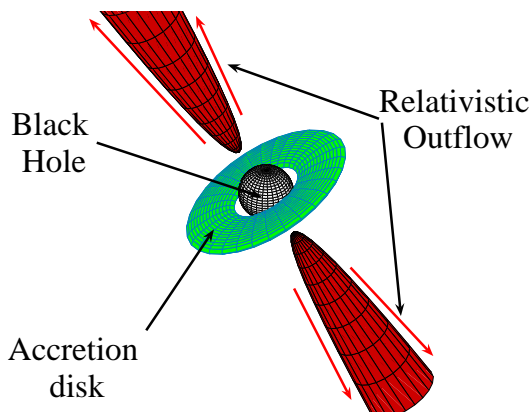}
\caption{GRBs and AGN are believed to be powered by black holes. The accretion of mass onto the black hole, through an accretion disk, releases large amounts of gravitational energy. If the black hole is rotating rapidly, another energy source becomes available: The rotational energy may be released by slowing the black hole down through interaction with the accretion disk. The energy released drives a jet-like relativistic outflow. The observed radiation is produced as part of the energy carried by the jets is converted, at large distance from the central black hole, to electromagnetic radiation.}
\label{fig:BHjet}
\end{figure}

The validity of the constraints imposed on the properties of candidate UHECR sources, as summarized in \S~\ref{ssec:source_constraints}, depends on the validity of the inference that the highest energy particles are protons, and on the validity of the assumption that the particles are accelerated by some electromagnetic process, for which the constraints derived in \S~\ref{ssec:power_gamma} are valid. The inference that the highest energy particles are protons is supported by the HiRes and PAO UHECR spectrum, by the properties of air showers as measured by HiRes, and by the anisotropy hints. However, the shower properties reported by PAO appear to be inconsistent with a pure proton composition at the highest energy (and possibly also with a heavy nuclei composition, see \S~\ref{ssec:primaries}). Given this, and the fact that the $pp$ cross section at the high energies under discussion is not well known, the possibility that the highest energy particles are heavy nuclei can not yet be excluded. If the particles are indeed heavy nuclei of charge $Z$, the minimum power requirement, Eq.~(\ref{eq:L}), would be reduced by a factor $Z^2$, and could possibly be satisfied by local steady sources like AGN (e.g. \cite{Peer09}).

Thus, although we have strong arguments suggesting that UHECR sources are protons produced by transient XG sources, and that the sources should satisfy the constraints given in \S~\ref{ssec:source_constraints}, which point towards GRBs being the likely sources, we are still missing a direct proof of the validity of these conclusions. The open questions that require conclusive answers are:
\begin{itemize}
    \item \textit{Composition.} Is the composition indeed dominated by protons, or is there a transition back to heavier nuclei at the highest energies? What is the cross section for $pp$ interaction at high, $>100$~TeV, energy?
    \item \emph{Galactic- XG transition.} At what energy does the flux become dominated by XG sources?
    \item \textit{Sources.} Are the sources indeed transient? If so, are the sources GRBs or other transients?
    \item \textit{Acceleration.} Are UHECRs accelerated, as suspected, in collisionless (relativistic) shocks? A theory of such shocks based on basic principles is not yet available (e.g. \cite{Waxman_rel-plasma_rev06} and refernces therein).
\end{itemize}

In addition to the open questions listed above, the physics of the candidate UHECR sources is also not well understood. As we have shown, UHECR sources are required to produce very large power and are likely to be driving relativistic outflows, see Eqs.~(\ref{eq:L}) and~(\ref{eq:Gamma}). These requirements suggest that the sources are powered by the accretion of mass onto black holes, as believed to be the case for GRBs and AGN. GRBs are most likely powered by the accretion of a fraction of a Solar mass on a $\sim1$~s time scale onto a newly born Solar mass black hole \cite{Meszaros02,Waxman03,Piran04,Meszaros06}. Recent observations strongly suggest that the formation of the black hole is associated with the collapse of the core of a very massive star. AGN are believed to be powered by accretion of mass at a rate of $\sim1$~Solar mass per year onto massive, million to billion Solar mass, black holes residing at the centers of distant galaxies \cite{Krolik98}. As illustrated in figure~\ref{fig:BHjet}, the gravitational energy released by the accretion of mass onto the black hole is assumed in both cases to drive a relativistic jet, which travels at nearly the speed of light and produces the observed radiation at a large distance away from the central black hole. The models describing the physics responsible for powering these objects, though successful in explaining most observations, are largely phenomenological: The mechanism by which the gravitational energy release is harnessed to drive jets, the mechanism of jet collimation and acceleration, and the process of particle acceleration (and radiation generation), are not understood from basic principles. In particular, the answer to the question of whether the jet energy outflow is predominantly electromagnetic or kinetic, which has major implications to our understanding of the mechanism by which the jets are formed, is not known despite many years of photon observations.
%This situation is common also to our understanding of Galactic micro-Quasars, which may be considered as a scaled down version of AGN \cite{Mirabel99}, with $\sim1$ Solar mass black hole (or neutron star) ``engines."

These open questions are unlikely to be answered by UHECR observatories alone. For example, given the uncertainties in the high energy $pp$ cross section, it is not clear that studying shower properties would determine the primary composition. The composition could be constrained by an energy dependent anisotropy study (see \S~\ref{ssec:aniso}). However, the conclusions of such an analysis would depend on some assumptions regarding the sources \cite{Lemoine09}. In addition, UHECR observatories are unlikely to identify the sources. Although they may provide a conclusive evidence for the correlation between the distribution of UHECR sources and that of matter in the local universe, and possibly discriminate between steady and transient sources (which may require large exposure that can be provided only by space-born detectors, see \S~\ref{ssec:predictions}), this would not determine which type of objects the sources are. It should be emphasized that electromagnetic observations are equally unlikely to resolve the open questions: despite many years of observations we are still lacking direct evidence for acceleration of nuclei in any astrophysical object, and fundamental questions related to the physics of the sources (e.g. the content of relativistic jets) remain unanswered.

Thus, resolving the UHECR puzzles would require a ``multi-messenger" approach, combining data from UHECR, $\gamma$-ray and neutrino detectors. Neutrino astronomy is likely to play an important role in this context: detection of GZK neutrinos (see \S~\ref{ssec:GZK}), combined with accurate measurements of the UHECR flux and spectrum, may allow us to determine the UHECR composition (and constrain the UHE $pp$ and neutrino interaction cross sections); detection of high energy neutrino emission from electromagnetically identified sources may allow us to identify the UHECR sources; Neutrino observations will provide new constraints on the physics driving the sources, which can not be obtained using electromagnetic observations, since they can escape from regions which are opaque to electromagnetic radiation (see \S~\ref{ssec:grb_nu} for examples related to GRBs).

Finally, it should be realized that if the UHECR sources are steady, identifying the sources by directly detecting their neutrino emission is highly improbable, due to the fact that the effective area of a 1~km$^2$ neutrino detector is $\approx3\times10^{-4}{\rm km^2}$ at $10^3$~TeV, $\approx 10^{-7}$ of the area of $>10^{19}$~eV CR detectors (hence, neutrinos will not be detected unless the neutrino luminosity of the sources exceeds their UHECR luminosity by a factor of $10^3$). In this case, identifying the sources will require a theoretical analysis combining electromagnetic, CR and neutrino data.

%\bibliography{uhecr}
%\bibliographystyle{hapj}
%\bibliographystyle{spmpsci}
%\bibliographystyle{spphys}

\end{document}